\documentstyle[12pt]{article}
\begin{document}

\author{Ary W. Espinosa--M\"uller \\
Departmento de F{\'\i}sica, Universidad de Concepci\'on \\
Casilla 4009, Concepci\'on, Chile \\
\and 
Adelio Matamala \\
Departamento de F{\'\i}sico--Qu{\'\i}mica, Universidad de Concepci\'on \\
Casilla 3--C, Concepci\'on, Chile}

\title{ALGEBRAIC FORMULATION OF THE OPERATORIAL PERTURBATION THEORY. PART 2. 
APLICATIONS}

\maketitle

\begin{abstract}

The algebraic approach to operator perturbation method has been applied to
two quantum--mechanical systems ``The Stark Effect in the Harmonic
Oscillator'' and ``The Generalized Zeeman Effect''. To that end, two
realizations of the superoperators involved in the formalism have been
carried out. The first of them has been based on the Heisenberg--Dirac
algebra of $\hat{a}^{\dagger }$, $\hat{a}$, $\hat{1}$ operators, the second
one has been based in the angular momemtum algebra of $\hat{L}_{+}$, $\hat{L}
_{-}$ and $\hat{L}_0$ operators. The successful results achieved in
predicting the discrete spectra of both systems have put in evidence the
reliability and accuracy of the theory.

\end{abstract}

\newpage

\section{Introduction}

The principal aim of this paper is to present how easily the algebraic
approach to perturbation methods developed in Part 1 of this series \cite
{Part 1} works.

The practice of ladder operators in the Operator Perturbation Method \cite
{de la pena} have led to general lemmas that make easier the applications of
the present {\bf A}{\em lgebraic {\bf F}ormulation of the {\bf O}perator 
{\bf P}erturbation Theory }({\bf AFOPT}) to non--relativistic quantum
mechanics. A fruitful procedure is to write the parallel projection
superoperator $\Pi $, the derivation superoperator $\Gamma $ and its inverse 
$\Gamma ^{-1}$ in terms of the ladder operators of a given algebra
associated with the physics of the problem at hands. This algebraic approach
to operator perturbation method can be applied, in principle, to any problem
provided that one can build the explicit and proper form of the ladder
operator common to $\hat{H}^0$ and the perturbation $\hat{V}$.

The paper is arranged as follows: in order to study the first system, i.e. a
charged harmonic oscillator the $\hat{a}^{\dagger }$, $\hat{a}$ and $\hat{1}$
, ladder operators of the Heisenberg--Dirac algebra has been introduced in
Sect. \ref{sec 2}. Then, we recover the $\Pi $, $\Gamma $ and $\Gamma ^{-1}$
superoperators in terms of the before mentioned ladder operators. To study
the second system, this time the Generalized Zeeman Effect, the spherical
base of angular momentum operators $\hat{L}_{+}$, $\hat{L}_{-}$ and $\hat{L}
_z$ have been introduced in Sect. \ref{sec 3}. Here we retrieve the
aforementioned set of superoperators, now in terms of the ladder operators
associated with the angular momentum algebra.

In Sect. \ref{sec 4} and \ref{sec 5} all the machinery developed in Part 1
of this series has been totally applied to both quantum mechanical systems.

In the last Sect. \ref{sec 6} the paper concludes with a general discussion
highlightening the reliability and accomplishment of the theory.

\section{The Stark Effect in the Harmonic Oscillator\label{sec 2}}

Let us consider a particle of charge $e$ and mass $m$, oscillating about its
equilibrium position, with fundamental frecuency $\omega _0$, subjects to a
homogeneous electric field of strength ${\cal E}$. We assume that the
oscillations are parallel to the direction of field.

The full Hamiltonian $\hat H$ is split into a zero order Hamiltonian $\hat
H^0$ and a perturbation $\hat V$

\[
\hat H=\hat H^0+\lambda \hat V\qquad \lambda \in \left[ 0,1\right] 
\]

Next we introduce the raising and lowering operators $\hat a^{\dagger }$ and 
$\hat a$ through the usual canonical transformation 

\[
\hat q=\sqrt{\frac \hbar {2m\omega _0}}\left( \hat a^{\dagger }+\hat a\right) 
\]

\[
\hat p=i\sqrt{\frac{\hbar m\omega _0}2}\left( \hat a^{\dagger }-\hat a\right) 
\]

with 

\[
\left[ \hat a^{\dagger },\hat a\right] =\hat 1 
\]

As it is well--known, the zero order Hamiltonian may now be written as 

\[
\hat H^0=\left( \hat a^{\dagger }\hat a+\frac 12\hat 1\right) \hbar\omega_0. 
\]

The perturbation operator is 

\[
\hat{V}=-e{\cal E}\hat{q}=-e{\cal E}\sqrt{\frac \hbar {2m\omega _0}}\left( 
\hat{a}^{\dagger }+\hat{a}\right) 
\]

Here $\hat{q}$ is the position operator of the particle regarding its
equilibrium position. The electric field strength plays the role of the
perturbation parameter $\lambda $. Hence, when the perturbation is switched
on, the zero order eigenkets$\left| n^0\right\rangle $ evolve into
orthogonal perturbed eigenkets $\left| n\right\rangle $ of energy 
$\varepsilon_n$. Then from Part 1 we may write 

\[
\left| n\right\rangle =\hat{U}^{\dagger }\left| n^0\right\rangle 
\]

and 

\[
\varepsilon_n=\varepsilon_n^0+\left\langle n^0\right|\hat W\left|n^0\right 
\rangle 
\]

Resolving the eigenvalue problem for the full Hamiltonian $\hat{H}$ involves
to find the unitary transformation $\hat{U}=\exp\left(\hat{G}\right)$, which 
must fulfill the requirement 

\[
\left[ \hat{H},\hat{W}\right] =\hat{0} 
\]

The unitary transformation will allow us to find the explicit form of the $
\hat{W}$ operator \cite{Part 1}.

From Part 1, it can be seen that three superoperators are of special
importance to algebraic formulation of the perturbation method. In fact, from 
the following equations 

\begin{equation}  \label{dosdies}
\hat W=\sum_n\hat W_n\quad ,\quad \hat W_n=\Pi \left( \hat A_n\right)
\end{equation}

and 

\begin{equation}  \label{dosonce}
\hat G=\sum_n\hat G_n\quad ,\quad \hat G_n=\Gamma ^{-1}\left( \hat A_n-\Pi
\left( \hat A_n\right) \right)
\end{equation}

it is apparent that $\Pi $, $\Gamma $ and $\Gamma ^{-1}$ are such
superoperators.

\subsection{Expressions for $\Pi$, $\Gamma$ and $\Gamma^{-1}$ superoperators 
in terms of $\hat{a}^{\dagger }$ and $\hat{a}$ ladder operators\label{sec 3}}

We look now at a representation of the superoperator $\Pi$, $\Gamma$ and 
$\Gamma^{-1}$ in terms of raising and lowering operators $\hat{a}^{\dagger}$
and $\hat{a}$. The straightforward application of the method outlined in
Sect. 3 of Part 1, allows us to write for the $\Pi $ superoperator, the
algebraic form 

\begin{equation}
\Pi \left( \hat{a}^{\dagger m}\hat{a}^n\right) =\delta _{mn}\hat{a}^{\dagger
m}\hat{a}^n  \label{tresuno}
\end{equation}

Analogously the same reference of Part 1, leads us to write for the $\Gamma $
superoperator the expression 

\[
\Gamma \left( \hat{a}^{\dagger m}\hat{a}^n\right) =\left( m-n\right) \hbar
\omega _0\,\hat{a}^{\dagger m}\hat{a}^n 
\]

Finally, from the well--defined $\Gamma ^{-1}$ superoperator we may write 

\begin{equation}
\Gamma ^{-1}\left( \hat{a}^{\dagger m}\hat{a}^n\right) =\left\{ 
\begin{array}{cc}
\hat{0} & {\rm if}m=n \\ 
&  \\ 
\left[ \left( m-n\right) \hbar \omega _0\,\right] ^{-1}\hat{a}^{\dagger m}
\hat{a}^n & {\rm if}m\neq n
\end{array}
\right.  \label{trestres}
\end{equation}

Notice that the integers $m$, $n$ do not label the energy level, really they
label the powers of $\hat a^{\dagger }$ and $\hat a$ in the multilinear
operator $\hat a^{\dagger m}\hat a^n$.

\subsection{Solution of the Perturbation Equations\label{sec 4}}

According to Eqs. \ref{dosdies} and \ref{dosonce} the solution of the
perturbation equations implies to find the $\hat{A}_n$ operators, which in
turn are determined through the commutator equation in terms of $\hat{H}^0$, 
$\hat{V}$ and the $\hat{G}_m$, $m<n$ operators.

From the mnemonic box diagrams given in the appendix of Part 1 the explicit
form of the commutator equations, follow straightfowardly.

\subsubsection{First Iteration}

\[
\hat{A}_1=(1)=\hat{V} 
\]

\begin{eqnarray*}
\hat{W}_1&=&\Pi \left( \hat{V}\right) \\
&=&-e{\cal E}\sqrt{\frac \hbar {2m\omega _0}}\Pi \left( \hat{a}^{\dagger }+
\hat{a}\right) \\
&=&-e{\cal E}\sqrt{\frac \hbar {2m\omega _0}}\left\{ \Pi \left( \hat{a}
^{\dagger }\right) +\Pi \left( \hat{a}\right) \right\}
\end{eqnarray*}

But according to Eq. \ref{tresuno} follows 

\[
\Pi\left(\hat{a}^\dagger\right)=\hat{0}\quad{\rm and}\quad\Pi\left( \hat{a} 
\right)=\hat{0} 
\]

Thus 

\[
\hat W_1=\Pi \left( \hat A_1\right) =\hat 0 
\]

Now we must calculate $\hat G_1$

\begin{eqnarray*}
\hat G_1&=&\Gamma ^{-1}\left( \hat A_1-\Pi \left( \hat A_1\right) \right) \\
&=&\Gamma ^{-1}\left( \hat A_1\right) \\
&=&-e{\cal E}\sqrt{\frac \hbar {2m\omega _0}}\Gamma ^{-1}\left( \hat
a^{\dagger }+\hat a\right)
\end{eqnarray*}

From Eq. \ref{trestres} we have 

\[
\Gamma ^{-1}\left( \hat{a}^{\dagger }\right) =\left( \hbar \omega
_0\,\right) ^{-1}\hat{a}^{\dagger } 
\]

and 

\[
\Gamma ^{-1}\left( \hat a\right) =\left( \hbar \omega _0\,\right) ^{-1}\hat a 
\]

Altogether, we may write 

\[
\hat G_1=-\frac{e{\cal E}}{\hbar \omega _0}\sqrt{\frac \hbar {2m\omega _0}}
\left( \hat a^{\dagger }+\hat a\right) 
\]

\subsubsection{Second Iteration}

\[
\hat{A}_2=(1\mid 1)\oplus (1,1\mid 0) 
\]

Having in mind the rules of box diagrams of Part 1 we may write down 

\[
\hat A_2=\frac 1{1!}\left[ \hat G_1,\hat V\right] +\frac 1{2!}\left[ \hat
G_1,\left[ \hat G_1,\hat H^0\right] \right] 
\]

The evaluation of this commutators is straightforward, thus we have 

\[
\left[ \hat G_1,\hat V\right] =-\frac{e^2{\cal E}^2}{m\omega _0^2}\hat 1 
\]

\[
\left[ \hat G_1,\left[ \hat G_1,\hat H^0\right] \right] =\frac{e^2{\cal E}^2
}{m\omega _0^2}\hat 1 
\]

Therefore 

\[
\hat A_2=-\frac 12\left( \frac{e^2{\cal E}^2}{m\omega _0^2}\right) \hat 1 
\]

Since $\Pi \left( {\rm cte}\right) ={\rm cte}$, we get 

\[
\hat W_2=\Pi \left( \hat A_2\right) =-\frac 12\left( \frac{e^2{\cal E}^2}{
m\omega _0^2}\right) \hat 1 
\]

and it is immediate that 

\[
\hat G_2=\Gamma ^{-1}\left( \hat A_2-\Pi \left( \hat A_2\right) \right)
=\hat 0 
\]

\subsubsection{Higher Iterations}

Proceeding with higher iterations it is easy to see that the $\hat A_3$, $
\hat A_4$, $\ldots $ operators are all null, since $\hat G_2=\hat 0$.

Finally, from Eqs. \ref{dosdies}, \ref{dosonce} we obtain 

\[
\hat{W}=-\frac 12m\left( \frac{e{\cal E}}{m\omega _0}\right) ^2\hat{1} 
\]

and 

\[
\hat G=\frac i\hbar \frac{e{\cal E}}{m\omega _0^2}\,\hat p 
\]

To sum up, the energies and the eigenvectors of the full Hamiltonian become 

\[
\varepsilon _n=\varepsilon _n^{\circ }-\frac 12m\left( \frac{e{\cal E}}{
m\omega _0}\right) ^2 
\]

and 

\[
\left| n\right\rangle =\exp \left( -\frac i\hbar \frac{e{\cal E}}{m\omega
_0^2}\,\hat p\right) \left| n^0\right\rangle 
\]

These results exactly coincide with those known from the elementary courses
of quantum mechanics. They show the reliability and accuracy of the
calculations achieved with our algebraic formulation of the operator
perturbation theory.

\section{Generalized Zeeman Effect\label{sec 5}}

To study this quantum mechanical system we will introduce the algebra of
angular momentum operator. But instead of working with cartesian component
operators $\hat L_x$, $\hat L_y$ and $\hat L_z$, it will be more convenient
for the algebraic formalism to use the spherical representation defined by 

\[
\hat L_{+}=\hat L_x+i\hat L_y 
\]

\[
\hat L_{-}=\hat L_x-i\hat L_y 
\]

\[
\hat L_0=\hat L_z 
\]

Where the $\hat L_{+}$ and $\hat L_{-}$ are the raising and lowering angular
momentum operators. For the sake of completeness we add the well--known
relation: $\hat L^2=\hat L_x^2+\hat L_y^2+\hat L_z^2$, which satisfies the
fundamental commutator 

\[
\left[ \hat L^2,\hat L_z\right] =\hat 0 
\]

where from the two distinguished relations follow 

\[
\hat L^2\left| lm\right\rangle =l(l+1)\hbar ^2\left| lm\right\rangle 
\]

and 

\[
\hat L_z\left| lm\right\rangle =m\hbar \left| lm\right\rangle . 
\]

The $\left| lm\right\rangle $ will define the zero--order orthonormal
eigenkets $\left| lm\right\rangle $, a base whereon to build the orthonormal
eigenkets $\overline{\left| lm\right\rangle }$.

Now we will inquire about the representation of $\Pi $, $\Gamma $ and 
$\Gamma^{-1}$ in terms of the angular spherical operators, $\hat L_{+}$, 
$\hat L_{-}$ and $\hat L_0$.

Prior to tackle this problem we want to state the next result, easily to
prove by mathematical induction the $0$-component operator $\hat{L}_0$ and
the multilinear operator $\hat{L}_{+}^m\hat{L}_0^p\hat{L}_{-}^n$, which
satisfies the following commutator relation 

\[
\left[ \hat{L}_z,\hat{L}_{+}^m\hat{L}_0^p\hat{L}_{-}^n\right] =\left(
m-n\right) \hbar \hat{L}_{+}^m\hat{L}_0^p\hat{L}_{-}^n 
\]

The system is a hydrogen atom exposed to two constant uniform magnetic
fields: one of them is parallel to the $z$--axis and the other lies in the $
(x,y)$--plane perpendicular to the $z$--axis.

The full Hamiltonian of such system is 

\[
\hat H=\hat H^0+\hat V 
\]

where the unperturbed Hamiltonian is 

\[
\hat H^0=\hat H_R^0+\frac \alpha {r^2}\hat L^2+\kappa \hat L_z 
\]

here $\hat H_R^0$ is a radial Hamiltonian, $\alpha $ and $\kappa $ are
positive constants. The perturbation operator is 

\[
\hat V=a\hat L_x+b\hat L_y 
\]

The real parameters $a$ and $b$ satisfy the normalization condition 

\[
a^2+b^2=1 
\]

Along the same line of reasoning outlined in Part 1, we state now the
following expressions for $\Pi $, $\Gamma $ and $\Gamma ^{-1}$
superoperators in terms of the spherical base of operators 

\begin{equation}  \label{nada}
\Pi \left( \hat L_{+}^m\hat L_0^p\hat L_{-}^n\right) =\delta _{mn}\hat
L_{+}^m\hat L_0^p\hat L_{-}^n
\end{equation}

\[
\Gamma \left( \hat L_{+}^m\hat L_0^p\hat L_{-}^n\right) =(m-n)\hbar \kappa
\,\hat L_{+}^m\hat L_0^p\hat L_{-}^n 
\]

Furthermore, since the superoperator $\Gamma ^{-1}$ is well--defined we get 

\[
\Gamma^{-1}\left(\hat L_{+}^m\,\hat L_0^p\,\hat L_{-}^n\right)=
\left\{\begin{array}{cc}
\hat 0 & {\rm if}\,m=n \\ 
&  \\ 
\left[ \left( m-n\right) \hbar \kappa \,\right] ^{-1}\hat L_{+}^m\hat
L_0^p\hat L_{-}^n & {\rm if}\,m\neq n
\end{array}
\right. 
\]

It is relevant to remark that the above realization is independent of any
degeneracy of the Hamiltonian \cite{Part 1} and the magnitude of the
perturbation is also immaterial \cite{Primas}.

It is convenient for further calculations to transform the perturbation
operator. Since: 

\[
\hat L_x=\frac 12\left( \hat L_{+}+\hat L_{-}\right) \quad {\rm and}\quad
\hat L_x=\frac i2\left( \hat L_{+}-\hat L_{-}\right) 
\]

We write 

\[
\hat V=\frac 12\left(a+ib\right)\hat L_{+}+\frac 12\left(a-ib\right)\hat L_- 
\]

with $u=a+ib$ we have 

\[
\hat V=\frac 12u\hat L_{+}+\frac 12u^{*}\hat L_{-} 
\]

and 

\[
\left| u\right| ^2=1. 
\]

\subsection{Solution of the Perturbation Equation}

The zero--order eigenenergies suitable to the $\hat H^0$ Hamiltonian are
given by 

\[
\varepsilon _{lm}^0=\varepsilon _R^0+\frac \alpha {r^2}l(l+1)\hbar ^2+m\hbar
\kappa . 
\]

The eigenenergies of the respective perturbed Hamiltonian are 

\[
\varepsilon_{lm}=\varepsilon_{lm}^0+\left\langle lm\right|\hat W\left|lm 
\right\rangle 
\]

and the orthonormal eigenkets are 

\[
\overline{\left|lm\right\rangle}=\exp\left(-\hat G\right)\left|lm\right 
\rangle 
\]

Also we note from Part 1 that 

\[
\hat W=\sum_n\hat W_n\qquad n=1,2,\ldots 
\]

and 

\[
\hat G=\sum_n\hat G_n\qquad n=1,2,\ldots 
\]

where 

\[
\hat W_n=\Pi \left( \hat A_n\right) 
\]

and 

\[
\hat G_n=\Gamma ^{-1}\left( \hat A_n-\Pi \left( \hat A_n\right) \right) 
\]

\subsubsection{First Iteration}

\[
\hat{A}_1=(1)=\hat{V}=\frac 12u\hat{L}_{+}+\frac 12u^{*}\hat{L}_{-} 
\]

then 

\[
\hat W_1=\Pi \left( \hat A_1\right) =\frac 12u\Pi \left( \hat L_{+}\right)
+\frac 12u^{*}\Pi \left( \hat L_{-}\right) 
\]

Therefore from Eq. \ref{nada} we conclude that 

\[
\hat{W}_1=\hat{0} 
\]

Furthermore, 

\begin{eqnarray*}
\hat G_1&=&\Gamma ^{-1}\left( \hat A_1\right) \\
&=&\frac 12u\left( \frac 1{\hbar \kappa }\hat L_{+}\right) +\frac
12u^{*}\left( -\frac 1{\hbar \kappa }\hat L_{-}\right) \\
&=&\frac 1{2\hbar \kappa }\left( u\hat L_{+}-u^{*}\hat L_{-}\right)
\end{eqnarray*}

\subsubsection{Second Iteration}

The diagrammatic expansion for the $\hat{A}_2$ operator is 

\[
\hat{A}_2=(1\mid 1)\oplus (1,1\mid 0) 
\]

From the aforementioned rules follow 

\[
\hat A_2=\frac 1{1!}\left[ \hat G_1,\hat V\right] +\frac 1{2!}\left[ \hat
G_1,\left[ \hat G_1,\hat H^0\right] \right] 
\]

From the operatorial expression of $\hat G_1$ and $\hat V$ it is easy to get 

\[
\left[ \hat G_1,\hat V\right] =\frac 1\kappa \hat L_0 
\]

For the second commutator it is profitable to remark that 

\[
\left[ \hat G_1,\hat H^0\right] =-\Gamma \left( \hat G_1\right) 
\]

Thus 

\[
\hat G_1=\Gamma ^{-1}\left( \hat A_1\right) =\Gamma ^{-1}\left(\hat V\right) 
\]

Then, left multiplying by $\Gamma $ one obtains 

\[
\hat{V}=\Gamma \left( \hat{G}_1\right) 
\]

In all we may write 

\[
\frac 1{2!}\left[ \hat G_1,\left[ \hat G_1,\hat H^0\right] \right] =-\frac
12\left[ \hat G_1,\hat V\right] =-\frac 1{2\kappa }\hat L_0 
\]

In short, the $\hat A_2$ operator is 

\[
\hat A_2=\frac 1\kappa\hat L_0-\frac 1{2\kappa}\hat L_0=\frac 1{2\kappa}\hat 
L_0 
\]

The $\hat G_2$ operator is 

\[
\hat G_2=\Gamma^{-1}\left(\hat A_2\right)=\frac 1{2\kappa}\Gamma^{-1}\left( 
\hat L_0\right) 
\]

However from Eq. \ref{trestres} $m=n=0$, then it follows that 

\[
\hat{G}_2=\hat{0} 
\]

Hence any commutator involving $\hat{G}_2$ must be vanished. Therefore it
can be concluded that in the third iteration the only non null commutator
leads to the equation 

\[
\hat{A}_3=\frac 1{2!}\left[\hat{G}_1,\left[ \hat{G}_1,\hat{H}^0\right]\right] 
+\frac 1{3!}\left[ \hat{G}_1,\left[ \hat{G}_1,\left[ \hat{G}_1,\hat{H}^0 
\right] \right] \right] 
\]

\subsubsection{Third Iteration}

A straightforward calculation gives 

\[
\hat{A}_3=-\frac 1{3\kappa ^2}\hat{V} 
\]

Here it is easy to check that 

\[
\hat W_3=\hat 0 
\]

and 

\[
\hat{G}_3=-\frac 1{3\kappa ^2}\hat{G}_1 
\]

\subsubsection{Fourth Iteration}

Alongside the same calculation we arrive to 

\[
\hat A_4=-\frac 1{8\kappa ^3}\hat L_0 
\]

Then 

\[
\hat W_4=-\frac 1{8\kappa ^3}\hat L_0 
\]

and 

\[
\hat G_4=\hat 0 
\]

From the expansion given by Eq. \ref{dosdies} and Eq. \ref{dosonce} , one
arrives to 

\[
\hat{G}=\left( \frac 1\kappa -\frac 1{3\kappa ^2}+\cdots \right) \left( u
\hat{L}_{+}-u^{*}\hat{L}_{-}\right) 
\]

and 

\[
\hat W=\left( \frac 1{2\kappa }-\frac 1{8\kappa ^3}+\cdots \right) \hat L_0 
\]

From the above results we can say that our algebraic calculations are
coincident with those early reported by Arthurs and Robinson \cite{Arthurs}.

\section{Summary and Discussions\label{sec 6}}

In order to show the subsequent stages involved in our calculations of the
discrete spectra of quantum mechanical systems, two systems have been
chosen: ``The Stark Effect in the Harmonic Oscillator'' and ``The
Generalized Zeeman Effect''. These systems have been studied with greater
details throughout so as to show how the present algebraic formalism works.

The sharp coincidence of the results so obtained confirms the reliability
and easiness of the calculations involved.

Another aspect of the method is the internal coherency and self consistency.
In other words it has not been necessary to introduce any unconnected
elements with the theory, as well as trials and errors assays \cite{Arthurs}
in order to carry out the calculations of the spectra of both systems.

Finally some practical consequences may be drawn from the performed
algebraic operations. Sometimes the commutator equations of higher orders
seem to be very involved and tiresome. However, just a bit of systhematic
work on this kind of equations reveals that many expressions have already
been calculated or an unknown quantity may be written in terms of another
quantity already calculated.

{\bf Acknowledgments}

We thank to Miss Paula J. Espinosa M. and Mrs A. H\'{a}sbun for subsequent
helps and for reading the manuscript. One of us (A.W.E.M.) specially thanks
Mr. Guillermo Rubilar A. for reading and for typing the paper.

This research was supported by FONDECYT grants 1989-0657.


\begin{thebibliography}{9}

\bibitem{Part 1} Part 1 to be published in The special issue of THEOCHEM
(1995).

\bibitem{de la pena} L. De la Pe\~{n}a and R. Montemayor, {\it Am. J. Phys.
} {\bf 48}, 855 (1980).

\bibitem{Primas} H. Primas, {\it Rev. Mod. Phys}. {\bf 35} (1963) 710; {\it 
Helv. Phys. Acta} {\bf 34}, 331 (1961).

\bibitem{Arthurs} A. M. Arthurs and P. D. Robinson, {\it Am. J. Phys.} {\bf 
37}, 921 (1969).

\end{thebibliography}
\end{document}